\documentstyle[aps, prbbib, epsf, twoside]{revtex} 
\begin{document} 

\draft 

\wideabs{
\title{
Investigation of Diamond Nucleation under Very Low Pressure in Chemical
Vapor Deposition.}

\author{Qijin Chen} 

\address{ State Key Laboratory of Surface Physics,
Institute of Physics, Chinese Academy of Sciences, Beijing, 100080, China}
\address{ Department of Physics, University of Chicago, 5720 S. Ellis
Ave., Chicago, IL $60637^\star$\\( $^\star$The corresponding address.
Email: qchen@rainbow.uchicago.edu)}
\date{\today} 
\maketitle


\begin{abstract}
	Diamond nucleation under very low pressure (0.1-1.0 torr) was
obtained at very high nucleation densities and very rapid rates using
hot-filament chemical vapor deposition (HFCVD). The density on
mirror-polished silicon was as high as $10^{10}$-$10^{11}$
$\mbox{cm}^{-2}$,
equivalent to the highest density in a microwave-plasma CVD system. That
on scratched silicon substrates was up to $10^9$ $\mbox{cm}^{-2}$, 1-2
orders of magnitude higher than that obtained under conventionally low
pressure (tens of torr, $10^7$-$10^8$ $\mbox{cm}^{-2}$). Also, the density
on scratched titanium substrates was as high as $10^{10}$
$\mbox{cm}^{-2}$. The samples were characterized using scanning electron
microscopy (SEM) and Raman spectroscopy. The mechanism is investigated in
detail, revealing that, under very low pressure, very long mean free path
of the gas species, strong electron emission from the hot filament, and
high efficiency of decomposition of hydrocarbon species by the filament
greatly increase the concentration of reactive hydrocarbon radicals and
atomic hydrogen on the substrate surface, and therefore, dramatically
enhance the nucleation eventually. This work has great practical
applications and theoretical significance. 
\end{abstract}

\pacs{81.15.Gh, 81.15.-Z, 81.10.-h, 81.10.Aj}
}

\section{INTRODUCTION}

	 Nucleation is the first step of diamond growth in chemical vapor
deposition (CVD) under low pressure. Great progress has been made in this
respect. In the early days of diamond research using CVD, diamond was
deposited using diamond seeds as nuclei,\cite{Angus1,Angus2} or directly
on a single-crystal diamond.\cite{Lander} In 1982, diamond seeds were no
longer needed to get diamond nuclei owing to the work of Matsumoto {\it et
al}.\cite{Matsumoto} However, the density was too low to lead to
continuous films. Later on, nucleation was found to be effectively
enhanced by scratching the substrate surface,\cite{Mitsuda} which has been
one of the major methods for nucleation enhancement since then. For the
most intensively studied silicon substrate, the nucleation density is up
to $10^7$-$10^8$ $\mbox{cm}^{-2}$, whereas it is no more than $10^4$
$\mbox{cm}^{-2}$ for unscratched, mirror-polished silicon. In addition,
various other pretreatment methods of the substrate have also been tried
and showed varying degrees of nucleation enhancement, such as
predeposition of non-diamond carbon including graphite,\cite{Angus3,Feng}
amorphous carbon,\cite{Angus3,Morrish} diamond-like carbon,\cite{Singh}
and even coating with pump oil.\cite{Morrish} However, all these methods
result in unoriented nucleation, and therefore, become invalid in
achieving epitaxy of diamond on hetero-substrates. The problem of
achieving high-density nucleation on mirror-polished single-crystal Si has
to be solved for this purpose. Breakthrough in this regard arose owing to
Jeng {\it et al\/} in microwave plasma CVD (MPCVD);\cite{Jeng} they
obtained local, oriented nucleation with a density up to $10^7$
$\mbox{cm}^{-2}$ by negatively biasing the substrate. In 1990, Yugo {\it
et al\/} obtained an (unoriented)  nucleation density up to
$10^9$-$10^{10}$
$\mbox{cm}^{-2}$ using a similar method.\cite{Yugo} Up to date, the
highest density achieved using this method is $10^{10}$-$10^{11}$
$\mbox{cm}^{-2}$, as reported by Stoner {\it et al}.\cite{Stoner1} In
1993, Jiang {\it et al\/} reported observation of oriented nucleation
using negative bias.\cite{Jiang} So far, high-density nucleation of
diamond on Si in a MPCVD system is no longer a problem. In an HFCVD system,
similar bias method had been tried on mirror-polished substrates for a
long time without success until recently. Using a similar negative bias in
their HFCVD system, Zhu {\it et al\/} reported that the nucleation
enhancement took place only at the edge of the sample.\cite{Zhu} Chen and
Lin achieved high density of oriented nucleation using an
electron-emission-enhancement (EEE) method with and without a bias to the
substrate, demonstrating an EEE mechanism.\cite{Chen1,Chen2,Chen3}

	On a Ti substrate using the scratching enhancement under normally
low pressure, it has been reported that nucleation takes place only after
a long carburization process with a very poor density,\cite{Park} which
usually leads to serious hydrogenation of the substrate. 

    We report in this paper another completely different nucleation method
and investigate its mechanism. Usually, the pressure for nucleation ranges
from tens of torr to above one hundred torr, the same as that for growth. 
Under this condition, high-density nucleation can not be obtained on
mirror-polished Si surface without the use of bias or EEE method. The
density on a scratched sample is only $10^7$-$10^8$ $\mbox{cm}^{-2}$. In
contrast, as reported below, under very low pressures (0.1-1.0 torr),
nucleation can be achieved on mirror-polish Si substrates with a density
as high as $10^{10}$-$10^{11}$ $\mbox{cm}^{-2}$, equivalent to the
highest value in MPCVD. For scratched substrates, the density is also 1-2
orders of magnitude higher than that under normal pressures. On a
scrtached Ti wafer , the density can also be as high as $10^{10}$
$\mbox{cm}^{-2}$. The nucleation progresses at a much higher speed with a
more uniform distribution across the sample. This method has its own
special merits, e.g., high uniformity and rapidity, in comparison with
other nucleation enhancing methods and has great significance in diamond
deposition. 

This work is mainly taken from Sec. 4.2 of the author's
thesis.\cite{Chen4} We did not notice the paper by Lee {\it et al\/}
until after its publication.\cite{Lee} However, part of the experimental
conditions and results were not properly reported, and the discussion was
very incomplete and only partly correct. In view of this and the
availability of ref.  \onlinecite{Chen4}, we deem it appropriate to
publish it on a more widely spread journal.  As the normal pressure
($10^1$-$10^2$ torr) is conventionally referred to low pressure, we call
the pressure (0.1-1.0 torr) in our experiments very low pressure, to show
the difference.

\section{EXPERIMENTS}

	Our experiments were carried out in an HFCVD apparatus as reported
in ref. \onlinecite{Chen5}. To repeat briefly, a $\phi 140$ mm and 500 mm long
fused silica tube was used as the deposition chamber. Coils of $\phi 0.2$ mm
tungsten wires were used as filaments. The substrates were Si and Ti
wafers in the size of $8\times10$ $\mbox{mm}^2$, mounted on a copper
platform under the filament during the experiments. The filament
temperature was measured with an optical pyrometer, while the substrate
temperature was measured with a thermocouple (Pt-PtRh).
 The source gas was methane diluted in hydrogen. Part of the substrates
were mirror-polished Si wafers without any nucleation enhancing
pretreatment, while the others were scratched with 0.5-1.0 $\mu\mbox{m}$
diamond powders. All samples were chemically cleaned with acetone in an
ultrasonic bath for $>10$ min, to remove oil, diamond residues after
scratching, and other dirties on the substrate surface before loaded into
the deposition chamber. In addition, the mirror-polished Si samples were
also cleaned subsequently with deionized water and rinsed in a 30 vol.~\%
HF solution to remove possible surface oxide. Typical experimental
parameters for nucleation are listed in Table~\ref{TableI}. The pressure
used, 0.1-1.0 torr, was one to two orders of magnitude lower as compared
with normal conditions.  Specific parameters of each experiment and the
growth conditions will be mentioned where appropriate. The as-deposited
samples were characterized with scanning electron microscopy (SEM) and
Raman spectroscopy. 

\begin{table}
\caption{Typical experimental conditions for nucleation.}
\begin{tabular}{llc}
Parameters              &       Notations       &values\\
\tableline
Flow rate (sccm\tablenotemark[1])       &       $F$     &       70-100\\
$\mbox{CH}_4$ concentration (vol.\%)&           [$\mbox{CH}_4$] &
1.0-3.0\\
Pressure (torr)         &       $p$     &       0.1-1.0\\
Filament temperature ($^\circ\mbox{C}$) &       $T_f$   &
2050-2150\\
Substrate temperature ($^\circ\mbox{C}$) & $T_s$        &
800-900\\
Nucleation time (min)   &       $t$     &       1-10
\end{tabular}

\tablenotetext[1]{sccm denotes cubic centimeter per minute at STP.}
\label{TableI}
\end{table}

\input epsf
\begin{figure} 
\centerline{
\leavevmode
\epsfbox{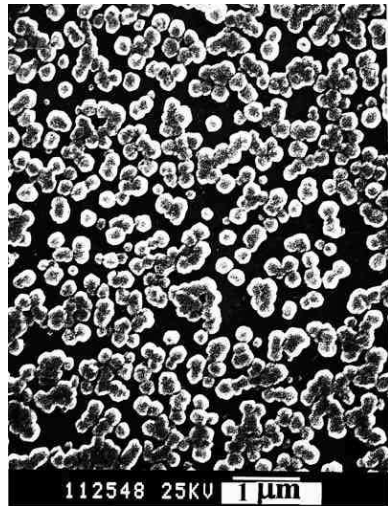}
}
\medskip

\caption{SEM image of the nuclei on a scratched Si(100)
substrate under very low pressure. $p = 1$ torr, [$\mbox{CH}_4$] = 1.5
vol.\%, $T_f = 2150^\circ \mbox{C}$, $T_s = 800^\circ \mbox{C}$, $F = 100$
sccm, $t = 10$ min. The density was $2\times 10^9$ $\mbox{cm}^{-2}$.}
\label{Fig.1} 
\end{figure}

\section{RESULTS}

	Figure \ref{Fig.1} shows the SEM image of the diamond nuclei on a
scratched
Si(100) substrate after 10 min nucleation under the conditions as follows. 
$p = 1$ torr, $F = 100$ sccm, $[\mbox{CH}_4] = 1.5$ vol.~\%, $T_f =
2150^\circ\mbox{C}$, $T_s = 800^\circ\mbox{C}$. The nucleation density was
measured to be $2\times10^9$ $\mbox{cm}^{-2}$, one to two orders higher than
the highest result ($10^7$-$10^8$ $\mbox{cm}^{-2}$) reported so far under
normal pressure (tens of torr). In addition, the nuclei were very
uniformly distributed, without any trace of nucleating along the
scratches.

\begin{figure}
\centerline{
\leavevmode
\epsfbox{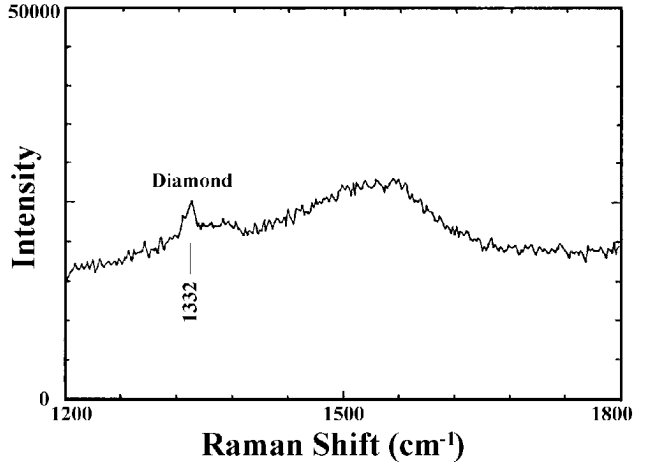}
}
\medskip
\caption{Raman spectrum of the sample shown in Fig.~\ref{Fig.1}. The
diamond signal at 1332 $\mbox{cm}^{-2}$ confirms the formation of diamond
nuclei.}
\label{Fig.2}
\end{figure}

	Figure \ref{Fig.2} shows the Raman spectrum of the sample in
Fig.~\ref{Fig.1}. The
characteristic peak of diamond at 1332 $\mbox{cm}^{-1}$ is very clear,
confirming the formation of diamond during 
\linebreak

\begin{figure}
\centerline{
\leavevmode
\epsfbox{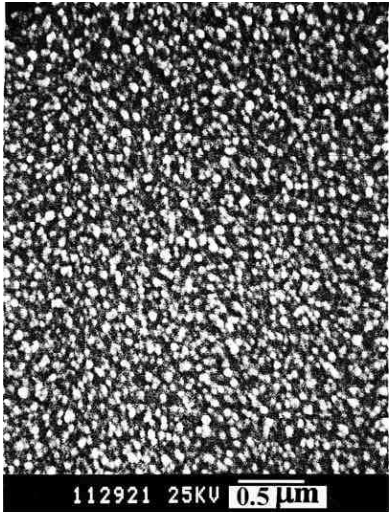}
}
\medskip

\caption{SEM image of a mirror-polished sample after 10 min
nucleation under
very low pressure. $p = 0.1$ torr, $F = 70$ sccm, [$\mbox{CH}_4$] = 2 vol.\%,
$T_f = 2150^\circ\mbox{C}$,
$T_s= 850^\circ\mbox{C}$. The density was as high as $10^{10}$-$10^{11}$
$\mbox{cm}^{-2}$. }
\label{Fig.3}
\end{figure}

\setlength{\parindent}{0 cm}
the nucleation. The broad band centered at $\sim 1540$ $\mbox{cm}^{-1}$ 
arises from non-diamond phase such as amorphous 
carbon, etc.  The diamond signal could not be very strong,
since a continuous diamond film had not formed as well
as the nuclei were very small.
\setlength{\parindent}{0.6 cm}

	As it turned out, it was not as easy to get high-density
nucleation on a mirror-polished substrate as on a scratched one under 1.0
torr. Thus the pressure was further lowered. Figure 3 shows the SEM photo
of a mirror-polished Si(100) wafer after 10 min nucleation under the
following conditions. $F = 70$ sccm, [$\mbox{CH}_4$] = 2 vol. \%, $p = 0.1$
torr, $T_f = 2150^\circ\mbox{C}$ and $T_s = 850^\circ\mbox{C}$. The
nucleation density was as high as $10^{10}$-$10^{11}$ $\mbox{cm}^{-2}$,
comparable to the highest density attained in MPCVD using negative bias
method.  Obviously, it was much higher than that on the scratched
substrate as shown in Fig.~\ref{Fig.1}.

\setlength{\parindent}{0.3 cm}
	As the nuclei in Fig. \ref{Fig.3} were too tiny, the layer of
nuclei 
\linebreak
\begin{figure}
\centerline{
\leavevmode
\epsfbox{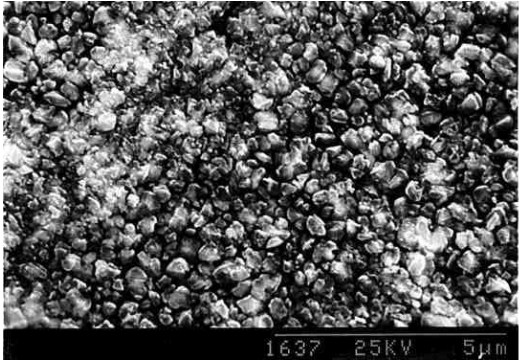}
}
\medskip
\caption{SEM picture of the diamond film on a mirror-polished
Si(100)
substrate after 5 min nucleation plus subsequent overnight growth. The
nucleation conditions were the same as in Fig.~\ref{Fig.3}. The growth
conditions
were: $F = 100$ sccm, [$\mbox{CH}_4$] = 0.7 vol.\%, $p = 20$ torr,
$T_f = 2000^\circ\mbox{C}$, $T_s =
800^\circ\mbox{C}$. }
\label{Fig.4}
\end{figure}

\begin{figure}
\centerline{
\leavevmode
\epsfbox{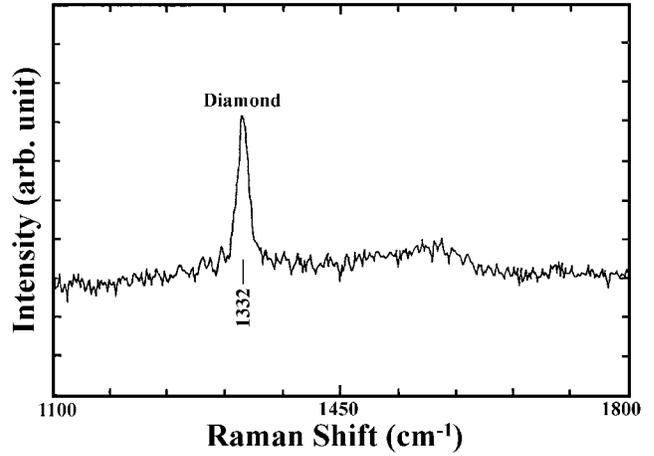}
}
\medskip
\caption{Raman spectrum of the sample in Fig.~\ref{Fig.4}.
The quality of the diamond film was pretty high. } 
\label{Fig.5}
\end{figure}

\setlength{\parindent}{0.0 cm}
was very thin. It is difficult to characterize that sample with Raman 
spectroscopy. To investigate whether 
diamond nuclei had indeed formed or
only a pure non-diamond carbon layer formed, subsequent growth was
tested. Figure \ref{Fig.4} shows the SEM image of a Si(100) sample after 5
min
nucleation under the same conditions as for Fig.~\ref{Fig.3} and a
subsequent
overnight growth at a normally low pressure. To eliminate amorphous carbon
in the nuclei more effectively, a lower $\mbox{CH}_4$ concentration was
used. The conditions for growth were as follows. $F = 100$ sccm,
[$\mbox{CH}_4$] = 0.7~vol.~\%,
 $p = 20$ torr, $T_f = 2000^\circ\mbox{C}$ and $T_s = 800^\circ
\mbox{C}$.  As is obvious in Fig.~\ref{Fig.4}, the diamond micro-crystals had
very good crystallinity with an average size of 0.5 $\mu\mbox{m}$ and a
density of $10^9$ $\mbox{cm}^{-2}$, which is comparable to the nucleation
density on the scratched substrate as shown in Fig.~\ref{Fig.1}. This
value was lower than the nucleation density in Fig.~\ref{Fig.3}, which 
can be attributed to the much larger size of the diamond 
crystallites than that of
the initial nuclei; a uniform size of 0.5 $\mu\mbox{m}$ would imply a
density of only $4\times10^8$ $\mbox{cm}^{-2}$.

	The Raman spectrum corresponding to Fig.~\ref{Fig.4} is 
\linebreak

\begin{figure}
\centerline{
\leavevmode
\epsfbox{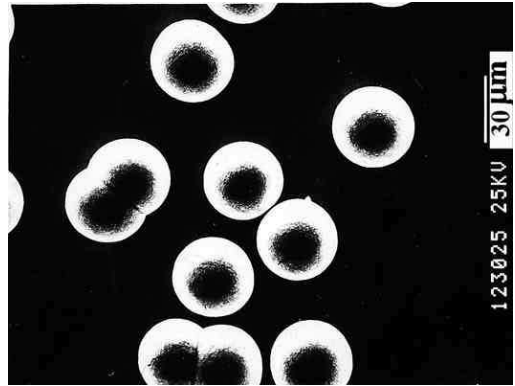}
}
\medskip
\caption{SEM image of nuclei on a mirror-polished Si substrate
after nucleation under normally low pressure. $p = 100$ torr,
[$\mbox{CH}_4$] = 0.7 vol.\%,
$F = 200$ sccm, $T_f = 2150^\circ\mbox{C}$, $T_s = 850^\circ\mbox{C}$.
The density was $3\times10^4$ $\mbox{cm}^{-2}$. }
\label{Fig.6}
\end{figure}

\begin{figure}
\centerline{
\leavevmode
\epsfbox{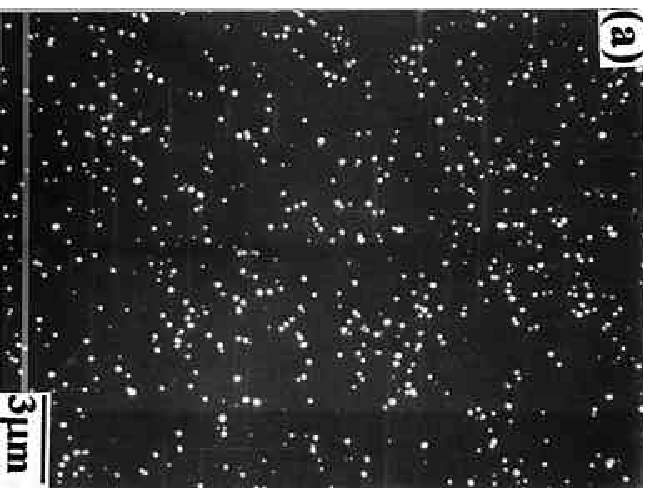}
}
\medskip
\centerline{
\leavevmode
\epsfbox{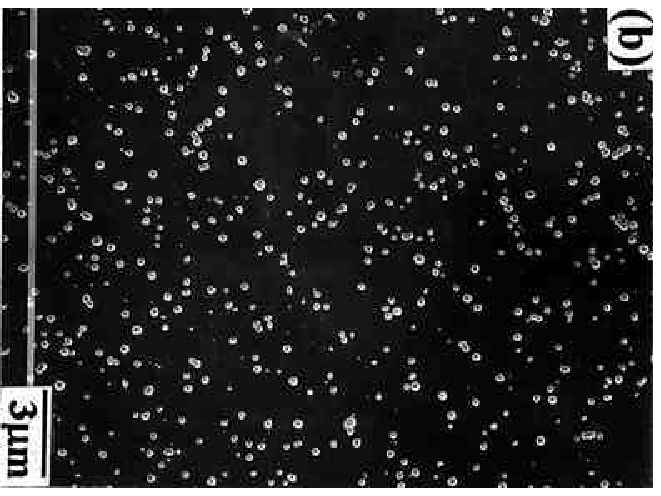}
}
\medskip
\centerline{
\leavevmode
\epsfbox{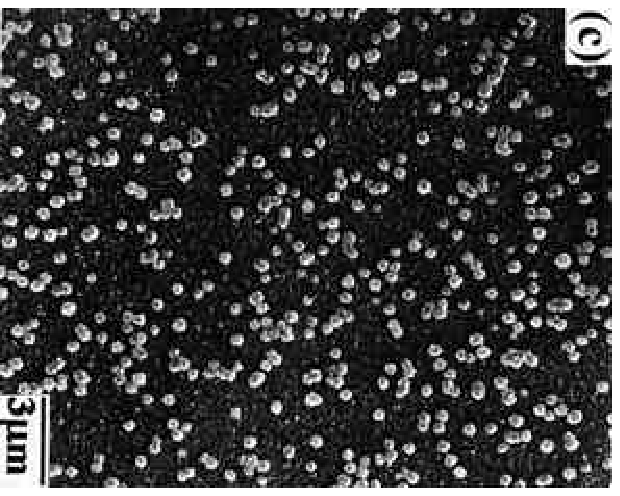}
}
\medskip
\centerline{
\leavevmode
\epsfbox{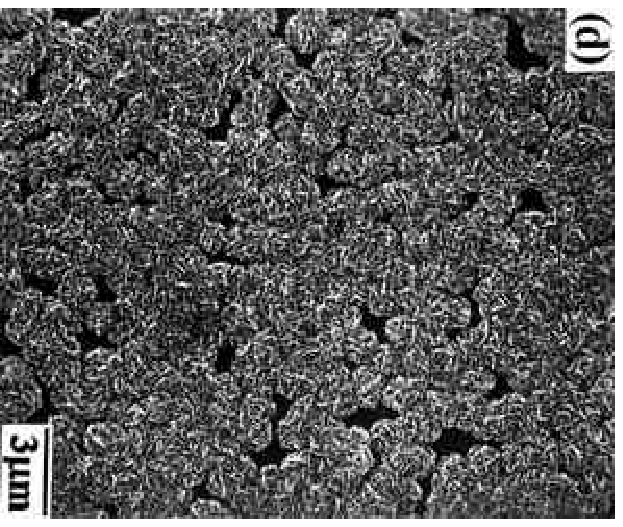}
}
\smallskip
\setlength{\parindent}{0 cm}

shown in Fig.~\ref{Fig.5}, demonstrating a very high quality of 
the diamond; the
diamond peak was very strong while that of non-diamond phase was very weak. 

\setlength{\parindent}{0.6 cm}
	Normally speaking, the nucleation density was only $10^4$
$\mbox{cm}^{-2}$ on a mirror-polished single-crystal Si substrate  without
pretreatment, as shown in Fig.~\ref{Fig.6}, which was 
\linebreak

\centerline{
\leavevmode
\epsfbox{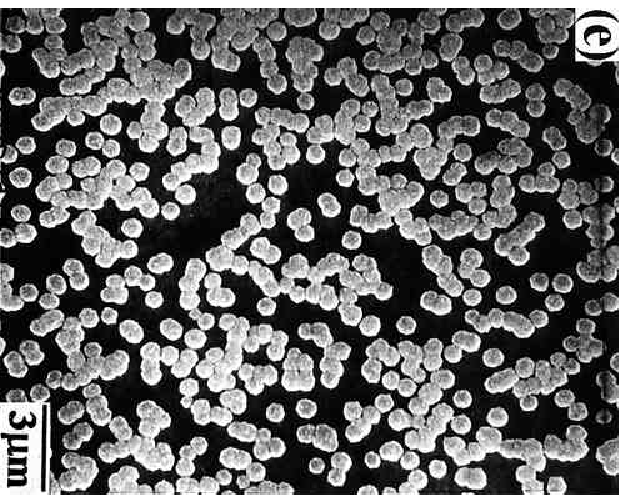}
}
\medskip
\centerline{
\leavevmode
\epsfbox{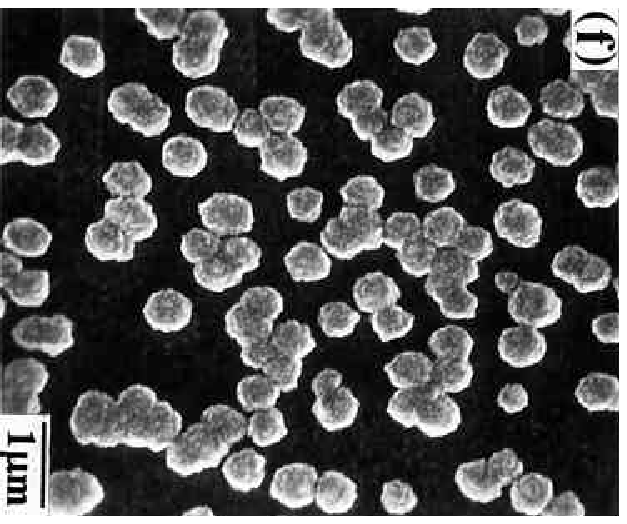}
}
\medskip

\caption{SEM images of scratched samples after nucleation under
very low
pressure (1 torr) for (a) 1 min, (b) 2.5 min, (c) 5 min; (d) 10 min, and
(e-f) 5 min plus 10 min subsequent growth under $p = 20$ torr.
Fig.~\ref{Fig.7}(f) is a
magnified image of (e). The density attained its final value
($3\times10^8$ $\mbox{cm}^{-2}$)
within the first minute. }
\label{Fig.7}
\end{figure}

\setlength{\parindent}{0 cm}

deposited under a
usual pressure of 100 torr. The nucleation density was only $3\times10^4$
$\mbox{cm}^{-2}$, several orders of magnitude lower than that for
Fig.~\ref{Fig.3}. 
\setlength{\parindent}{0.6 cm}

	In brief, on a scratched Si substrate, the density of $10^9$
$\mbox{cm}^{-2}$ was obtained under the very low pressure as compared with
the one of only $10^7$-$10^8$ $\mbox{cm}^{-2}$ under the normally low
pressure. On an unscratched, mirror-polished Si substrate,
$10^{10}$-$10^{11}$ $\mbox{cm}^{-2}$ was achieved in contrast with $10^4$
$\mbox{cm}^{-2}$ for the normal pressure. This demonstrates that using
very low pressure is a very effective method for high-density nucleation
on both scratched and unscratched substrates. 

	To demonstrate more clearly the nucleation rate, we studied the
development of nucleation process with the nucleation time. Figure
\ref{Fig.7}(a)-(d)  show the SEM image of the nuclei on scratched
substrates under the same nucleation parameters but with different nucleation 
time of 1 min, 2.5 min, 5 min and 10 min. To ensure the uniformity
of the scratches on the samples, all of them were cleaved from a large Si
wafer after scratching.  The nucleation parameters were: $F = 70$ sccm,
[$\mbox{CH}_4$] = 3~vol.~\%, $p = 1$ torr, $T_f =
2100$-$2150^\circ\mbox{C}$, and $T_s = 850$-$900^\circ\mbox{C}$.
  Since the time was very short,
it was difficult to adjust the filament and substrate temperatures to be
exactly the same.  Nonetheless, Fig.~\ref{Fig.7}(a)-(d) show consistently
\linebreak
\begin{figure}
\centerline{
\leavevmode
\epsfbox{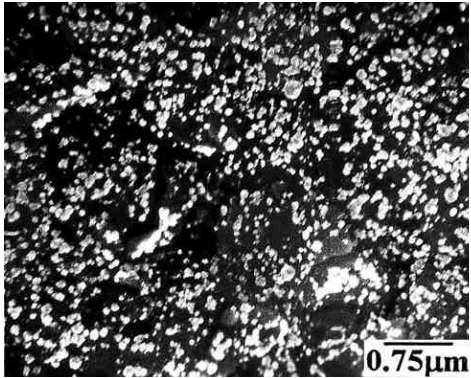}
}
\medskip
\caption{SEM image of the nuclei on a scratched Ti substrate after
2 min
nucleation under very low pressure ($p = 1$ torr). The density was as high
as $1.5\times10^{10}$ $\mbox{cm}^{-2}$. Nucleation conditions: $F = 70$ sccm,
[$\mbox{CH}_4$] = 3 vol.\%, $T_f
= 2050^\circ\mbox{C}$, $T_s = 850^\circ\mbox{C}$. }
\label{Fig.8}
\end{figure}

\setlength{\parindent}{0 cm}
 
progress of the nucleation process. The size of the nuclei
 grew rapidly, while the nucleation density, which was approximately
$3\times10^8$ $\mbox{cm}^{-2}$, almost attained its final value during the
first minute without much increase later on. Fig.~\ref{Fig.7}(e) shows a
sample after 5 min nucleation under the same condition plus 10 min 
subsequent growth under normal pressure. The growth parameters were: $F = 100$
sccm, [$\mbox{CH}_4$] = 1.5 vol.~\%, $p = 20$ torr, $T_f =
2050^\circ \mbox{C}$, and $T_s = 800^\circ \mbox{C}$. A magnified SEM picture
reveals that the crystalline shape was beginning to be clear, as shown in
Fig.~\ref{Fig.7}(f). The size of the diamond particles was larger that
that before growth (Fig.~\ref{Fig.7}(c)), but much smaller than that for
10 min nucleation only (Fig.~\ref{Fig.7}(d)), implying that the deposition
rate of carbon, including diamond and non-diamond, was much higher under the
low-pressure nucleation conditions above. This was one of the reasons why
the nucleation progressed so rapidly, as will be discussed below. 

\setlength{\parindent}{0.6 cm}
	Apart from Si substrates, very-low-pressure method was also
applied to other substrates. Figure 8 shows the SEM image of nuclei on a
scratched polycrystalline Ti substrate after 2 min nucleation under very
low pressure ($p = 1$ torr). The density was measured to be $1.5\times
10^{10}$ $\mbox{cm}^{-2}$.  This is an amazingly high density, and also
indicates and a very rapid nucleation rate, as compared with the report of
Park and Lee.\cite{Park} They reported that nucleation began only after an
intermediate TiC layer grew to as thick as 50 $\mu\mbox{m}$, with a very
poor density, which was lower than that on a scratched Si substrate under
normal pressure.  Therefore, the very low pressure method works not just
for Si substrates.

\section{DISCUSSIONS}

	As demonstrated clearly above, the very low pressure was
responsible for the nucleation enhancement. To make full use of this
method, a complete understanding of its mechanism is necessary. Generally
speaking, the carbon ad-atoms may diffuse into the substrates at high
temperatures while the substrate atoms may diffuse out. To nucleate, it is
critical to generate supersaturation of carbon atoms/radicals on the
substrate surface.\cite{Angus1} Upon saturation of carbon and/or
hydrocarbon species, graphite, amorphous carbon and diamond particles
begin to form, depending on experimental conditions. On the other hand,
the role of atomic hydrogen is also
critical.\cite{Spitsyn,Frenklach,Spear,Sun1} A sufficient amount of atomic
hydrogen is necessary to (i) extract the H atoms from the substrate
surface to create active nucleating sites, to (ii) suppress the formation
of carbon in $sp^2$ phase (i.e., non-diamond carbon) to ensure the
formation of carbon in $sp^3$ phase (i.e., diamond), and to (iii) help
eliminate possible oxide layers on the substrate surface, which is usually
regarded as a hindrance against nucleation. The reason that only very poor
nucleation is obtained on a mirror-polished substrate under normal
conditions is mainly because of (i) the lack of enough nucleating sites
and (ii) the low concentration of reactive hydrocarbon radicals and atomic
hydrogen.  Therefore, to get high density nucleation, enough amount of
nucleating sites have to be created first, and then sufficiently large
concentrations of reactive hydrocarbon radicals and atomic hydrogen have
to be provided.  For a Ti substrate, the easy formation of a TiC layer
usually precedes the formation of diamond nuclei. This makes the
supersaturation of carbon and/or hydrocarbon species on the substrate
surface even more important.  Based on these observations, several factors
help explain the nucleation enhancing effect of very low pressure, as
follows. 

	First, under very low pressure (0.1-1.0 torr), the mean free path,
$\lambda$, of the molecules and radicals of the source gas is 1-2 orders
of magnitude higher than that under normal pressure (tens of torr), as the
mean free path is in inverse proportion to the pressure at equilibrium. 
Furthermore, the probability for a molecule to move a distance $x$ without
collision is $e^{-x/\lambda}$, which is an exponential relationship. The
concentrations of reactive hydrocarbon radicals and of atomic hydrogen
decrease exponentially with increasing transportation distance. The
molecules are decomposed in the neighborhood of the hot filament, whereas
the nucleation and deposition takes place at a distance. Part of the
decomposed radicals recombine through collisions, which is not favorable
to deposition. The mean free path can be estimated as follows. The
temperature is not uniform between the filament (2300-2400K) and substrate
(1100K). For an estimate, let us take $T=1700$K as an average. Since quite
different sizes, and thus the mean free paths also differ much. Take
$r_0=1.0$ \AA\ as the radius of the background gas species in average
since the gas is mainly composed of molecular hydrogen.\cite{size,sqrt2} Then we
have 

 $\lambda = \frac{k_BT}{\sqrt{2}\pi \overline{d}^2p}, \qquad 
where \quad \overline{d} = r+r_0$\\ 
For hydrogen atoms, $r = 0.75$ \AA, $\lambda_{\mbox{H}} = 1.3/p$ mm; for
hydrocarbon radicals, mainly $\mathrm{CH_x}$, take $r = 1.6$ \AA, then
$\mathrm{\lambda_{CH_x}}$ = $0.59/p$ mm, where $p$ is in unit torr, and
$k_B$ is the Boltzman constant.  For normal pressure, say, $p = 20$ torr,
$\mathrm{\lambda_H} = 0.065$ mm, $\mathrm{\lambda_{CH_x}} = 0.03$ mm; for
$p = 1$ torr, $\mathrm{\lambda_H} = 1.3$ mm, $\mathrm{\lambda_{CH_x}} =
0.59$ mm, while for $p = 0.1$ torr, $\mathrm{\lambda_H} = 13$ mm,
$\mathrm{\lambda_{CH_x}} = 5.9$ mm. Take $x = 5$ mm, which was the
typical filament-substrate distance in our experiments.  So the
probabilities for a reactive hydrocarbon precursor to get to the sample
without collisions are $e^{-x/\mathrm{\lambda_{CH_x}}}\approx 4\times
10^{-73}$ (= 0!), $2\times10^{-4}$, and 0.43 for $p = 20$ torr, 1 torr,
and 0.1 torr, respectively. For atomic hydrogen, the probabilities are:
$e^{-x/\mathrm{\lambda_H}}\approx 4\times10^{-34}$, 0.02, and 0.68 for $p
= 20$ torr, 1 torr, and 0.1 torr, respectively. This is an amazingly
enormous difference. Therefore, the probability for the atomic hydrogen
and reactive hydrocarbon radicals in the neighborhood of the hot filament
to get directly onto the substrate surface without collision or
recombination is dramatically increased under very low pressure by many
orders of magnitude. On the other hand, as the flow rate of the source gas
did not change with pressure in our experiments, the amount of atomic
hydrogen and hydrocarbon radicals generated by the hot filament per unit
time remained
unchanged. Actually,
the amount increased due to the effect mentioned as the third reason
below. In result, the
concentrations of atomic hydrogen and of reactive hydrocarbon species were
greatly increased, leading to very high supersaturation of the
hydrocarbon. Since the nucleation rate is roughly proportional to the
amount of the reactive hydrocarbon species arriving at the substrate
surface per unit time, both the density and the rate of nucleation were
dramatically enhanced. 

	While the damages of the substrate surface resulting from
scratching may serve as nucleating sites, there exist only a very small
density of surface defects on the unscratched, mirror-polished substrates.
Therefore, a much lower pressure (0.1 torr) was used in our experiments
to get a much higher concentration of reactive hydrocarbon radicals and
hydrogen atoms to create an enough density of nucleating sites, and
finally an enough high nucleation density, based on the mean-free-path
scenario. On the other hand, as nucleation preferentially occurs on the
damages of a scratched sample surface, these damages may put an upper
limit of the density of nucleation. There seems to be no such an upper
limit for an unscratched sample. This is probably the reason why
 a much higher density was obtained on the mirror-polished substrate. 

	Next, very low pressure induced very strong electron emission from
the hot filament with an emission current as high as 0.5-1.0A. The
electrons had a continuous energy distribution from 0 up to $\sim$80 eV. 
As discovered in earlier work,\cite{Chen2,Chen3} the emitted electrons
collided with and disassociated various gas molecules or radicals, helped
to generate more atomic hydrogen and hydrocarbon radicals, and increased
the concentration of atomic hydrogen and hydrocarbon radicals near the
sample. 

	Last, under lower pressure, the efficiency of decomposition of the
gaseous species by the hot filament was higher. As reported by
Setaka,\cite{Setaka} lower pressure leads to a higher decomposition
efficiency of hydrogen. For methane, a similar result is expected. This
effect again helped to increase the concentration of atomic hydrogen and
reactive hydrocarbon radicals on the substrate surface. 

	All of these factors account for the nucleation enhancing effect
of the very low pressure in our experiments. Katoh {\it et al\/}
studied the influence of the pressure on the bias enhanced nucleation in
MPCVD, and reported that lower pressure led to a higher nucleation
density, in agreement with our results.\cite{Katoh} 

	Calculations have also been done by Spear {\it et al\/} on the
fraction of carbon deposited as a function of pressure for a mixture of
$\mathrm{CH_4}$-$\mathrm{H_2}$ at equilibrium.\cite{Spear} The result
shows that
fraction becomes higher under lower pressure while other conditions are
the same, consistent with our argumentation above.

	In comparison with the negative-bias enhancement method and/or the
EEE method, either in MPCVD or HFCVD, the very low pressure is
distinguished from its high uniformity and rapidity. For both the bias and
EEE methods, nucleation usually takes place
non-uniformly.\cite{Zhu,Chen3,Stoner2} Careful parameter control and
enough long time are necessary to get a uniform sample. As shown in
Fig.~\ref{Fig.7}(a)-(f), nucleation with a highly uniform distribution
occurred from the very beginning. 

	In the above calculations, we have assumed implicitly a
thermodynamical equilibrium of the gases in the deposition chamber, which
was not the situation in practice.  There existed a compulsive flow due to
the pump, which can be estimated using the equation of state of an ideal
gas, $pV = nRT$, where $V$ is the gas flow rate in unit of
$\mathrm{m^3/sec}$, n is the flow rate in unit of mole/sec, $T$ is the
temperature in Kelvin, and $R = 8.314$ J/(mole.K) is a constant. Let $v$
denote the flow velocity of the gas in unit of m/sec, then $V =
\frac{1}{4}\pi D^2v$, where $D = 0.140$ m is the diameter of the chamber.
Therefore, we have

$v = \frac{4nRT}{D^2p}$. \\
As $T$ was not uniformly distributed in the chamber, it is appropriate to
take cross-section between the filament and the substrate of the chamber
as a better estimate, which was approximately 1000K. Take the flow rate to
be 100 sccm, i.e., $n = 100$ sccm = $7.44\times10^{-5}$ mole/sec, and plug
in all the numbers, we get $v = 0.3/p$ m/sec, where $p$ is in unit torr.
This velocity is negligibly small even at $p = 0.1$ torr (3 m/sec), as
compared to the thermal velocity of the gas molecules, which is of the
order $10^3$ m/sec.  As a matter of fact, the cross-section of the inlet
of the source gases, which was at a distance of only several centimeters
away from the filament, was much smaller than that of the whole chamber,
the flow velocity between the filament and the substrate might be one
order of magnitude higher. Even so, it was still only a first order
perturbation to the thermal velocity. Accordingly, we expect that our
estimate about the mean free path remains valid. 

	On the other hand, as the filament was close to the inlet of the
gases, this compulsive flow velocity at very low pressure was very
important in that it guaranteed that the amount of the reactive
hydrocarbon radicals and atomic hydrogen getting to the substrate surface
per unit time was proportional to the gas flow rate in this steady,
non-equilibrium case, as opposed to being proportional to the pressure in
a static, equilibrium case, which was a good approximation only for the
case of normally low pressure. In the latter case, $v$ was really small
so that the amount of hydrogen and methane decomposed by the filament was
mainly proportional to the pressure, whereas in the former case, it was
mainly proportional to flow rate. While both factors were present, it is
believed that the former case dominated under the very low pressure in the
experiments.

	While the compulsive flow velocity did not make much difference in
terms of the velocity distribution and the mean free path of the gas
species, the energy distribution between the filament and the substrate
might be far from equilibrium under very low pressure. As shown above,
under normal pressure, the mean free path was so small, equilibrium did
arise. On the contrary, under very low pressure, say, 0.1 torr, 40\% of
hydrocarbon and 70\% atomic hydrogen could get onto the substrate without
collisions, while they still had the energy from filament at T = 2400 K,
more than twice that of the substrate, 1100K. Higher energy was believed
to be able to enhance the mobility and reactivity of the ad-species on the
substrate, and might therefore increase the nucleation rate.

	One may have noticed that, under normal pressure for a scratched
substrate, a considerably high density of nucleation is usually obtained
regardless of the very small mean free path, as shown above. The reasons
may be explained as follows. First, as most
 of the gas species are $\mbox{H}_2$ molecules, they do not likely result
in loss of atomic hydrogen or hydrocarbon radicals in collisions. Only
collisions between hydrogen atoms and/or hydrocarbon radicals may result
in recombination and counter the effect of decomposition by the filament. 
The effective ``mean free path'' in terms of recombination will be larger. 
Thus, the probabilities for a hydrogen atom and a reactive hydrocarbon
radical to get to the substrate surface are much higher than those
calculated above,
 although they may undergo collisions. Second, Apart from $\mathrm{CH_x}$
($\mbox{x}<4$) and $\mathrm{C_2H_x}$, $\mbox{CH}_4$ also contributes in
nucleation,\cite{Sun2} whose concentration is independent of the mean free
path. It can occupy a vacant surface site, and help to form nuclei, though
its role is much less important as compared with $\mathrm{CH_x}$
($\mbox{x}<4$) and $\mathrm{C_2H_x}$, etc.  Third, part of the
decomposition may take place in the neighborhood of the substrate, though
the decomposition rate is very low, as the substrate temperature is far
from high enough. While these contributions may lead to considerable
nucleation density on a scratched substrate (at a low rate, though), they
are not sufficient to lead to a high density on an unscratched,
mirror-polished substrate. 

	As the substrate is usually covered with a surface oxide layer,
which is disadvantageous to high-density nucleation, we see that, under
very low pressure, a very high percentage of atomic hydrogen is
transported onto the substrate from near the filament without collisions.
This greatly helps to eliminate the surface oxide to make a clean
substrate surface, presenting a nucleation enhancing effect.

	While our very low pressure led to impressive nucleation enhancement,
however, we do not claim that the pressure should be arbitrarily
low. As the very low pressure was realized by increasing the pump rate
and/or decreasing the flow rate, the pressure could not be arbitrarily
low. First, the capability of the pump was limited; it is difficult to
get a high vacuum in the presence of gas feed. Second, at very low pressure,
the deposition
rate was roughly proportional to the flow rate, which should not be too low.
Third, the mean free path should not be too long in the neighborhood of
the filament, else the gas could not be effectively decomposed by the
filament, as most gaseous species might not pass through the hot filament zone. 
Last, under too lower pressure, non-equilibrium effect would be
more important, and the analysis above should be modified, accordingly.
Therefore, there exists an optimum pressure which gives rise to the best
result. Further experiments are necessary to optimize the experimental
parameters.

	It is appropriate to point out that the local pressure between the
filament and the substrate was a little higher than the average of the
whole chamber, however, the main conclusions, e.g., the much longer mean
free path under the much lower pressure, etc, should remain valid.

	Using the very low pressure technique, epitaxial nucleation on
Si(111) substrates had been achieved,\cite{Chen6} and the serious problem
of the formation of thick intermediate TiC layers and hydrogenation of
very thin Ti substrates resulting from the long, slow, poor nucleation
process has been solved.\cite{Chen5} It also sheds light on the mechanism
of diamond nucleation on a hetero-substrate, as it confirms that a high
concentration of reactive hydrocarbon species and atomic hydrogen is
necessary for nucleation. 

\section{SUMMARY}

	In summary, under very low pressure (0.1-1 torr), high density
nucleation was achieved on mirror-polished Si substrates with a density as
high as $10^{10}$-$10^{11}$ $\mbox{cm}^{-2}$, comparable to the highest
value for MPCVD. For scratched Si substrates, the density was as high as
$10^9$ $\mbox{cm}^{-2}$, 1-2 orders of magnitude higher than that obtained
under normal pressure (tens of torr). In addition to Si substrates, a
density of as high as $10^{10}$ $\mbox{cm}^{-2}$ was obtained on
polycrystalline Ti substrates. Detailed study of the process of nucleation
with increasing time revealed that nucleation progressed at a very high
speed, the density getting to its final value within a very short time (1
min).  Calculations demonstrate that, under very low pressure, the mean
free path of the gas molecules and/or radicals is increased by 1-2 orders
of magnitude, as a result, the probability for an hydrogen atom or an
reactive hydrocarbon radical to transport from the filament to the
substrate without collisions is dramatically, exponentially increased. On
the other hand, very low pressure induces very strong electron emission
with the electron energy up to 80 eV, adding to the disassociation of the
gas species. Moreover, very low pressure favors a high efficiency for the
filament to decompose the gas species. All these factors result in much
higher concentrations of the atomic hydrogen and reactive hydrocarbon
radicals on the substrate surface, leading to the drastically enhanced
nucleation.  This method solves the problem of getting high density of
diamond nucleation on mirror-polished substrates in addition to the EEE
method.  It has great practical applications and theoretical significance.
It is hoped that the very-low-pressure method can be extended for the
growth stage, not just for nucleation. More work still needs to be done
 in this regard. 

\acknowledgements

	The author thanks Z. Lin for various support as an advisor
while this work was done, Y. Chen for collaboration on the very beginning
of this work, Y.J. Yan , Q.L. Wu, and X. Kuang for operating the SEM, K.
Zhu and Q. Zhou for their Raman spectroscopy service, and MRSEC at the
University of Chicago for its computer facilities. This work is
financially supported by Chinese Natural Science Foundation, 863 Program
and Beijing Zhongguancun Associated Center of Analysis and Measurement.


\begin{references}
\bibitem{Angus1}J. C. Angus, Y. Wang, and M. Sunkara, Annu. Rev. Mater.
Sci. {\bf 21}, 221 (1991); 
  J. C. Angus, C. C. Hayman, Science {\bf 241}, 913(1988)
\bibitem{Angus2}J. C. Angus, H. A. Will and W. S. Stanko, J. Appl. Phys.
{\bf 39}, 2915(1968);  
  D. J. Poferl, N. C. Gardner and J. C. Angus, J. Appl. Phys. {\bf 44},
1428(1973);  
  S. P. Chauhan, J. C. Angus and N. C. Gardner, J. Appl. Phys. {\bf 47},
4746(1976)
\bibitem{Lander}J. J. Lander and J. Morrison, Surf. Sci. {\bf 2},
553(1964); J. Chem. Phys. {\bf 34}, 1403(1963); 
  B. V. Spitsyn, L. L. Bouilov, and B. V. Deryagin, J. Cryst. Growth {\bf
52}, 219(1981)  
\bibitem{Matsumoto}S. Matsumoto, Y. Sato, M. Kamo and N. Setaka, Jpn. J.
Appl. Phys. {\bf 21}, Part 2, 183(1982);
  S. Matsumoto, Y. Sato, M. Tsutsumi and N. Setaka, J. Mater. Sci. {\bf
17}, 3106(1982)
\bibitem{Mitsuda}K. Mitsuda, Y. Kojima, T. Yoshida and K. Akashi, J.
Mater. Sci. {\bf 22}, 1557 (1987)
\bibitem{Angus3}J. C. Angus, Z. Li, M. Sunkara, R. Gat, A. B. Anderson,
S. P. Mehandru and M. W. Geis, in {\it Proceedings of the
  2nd International Symposium on Diamond Materials\/}, edited by A. J. 
Purdes, J. C. Angus, R. F. Davis, B. M. 
  Meyerson, K. E. Spear, and M. Yoder (The Electrochemical Society,
Pennington, NJ, 1991), p.125
\bibitem{Feng}Z. Feng, K. Komvopoulos, I. G. Brown and D. G. Bogy, J.
Appl. Phys. {\bf 74}, 2841(1993)
\bibitem{Morrish}A. A. Morrish and P. E. Pehesson, Appl. Phys. Lett. {\bf
59}, 417(1991)
\bibitem{Singh}J. Singh and M. Vellaikal, J. Appl. Phys. {\bf 73},
2831(1993); 
  K. V. Ravi and C. A. Koch, Appl. Phys. Lett. {\bf 57}, 348(1992); 
  K. V. Ravi, C. A. Koch, H. S. Hu and A. Joshi, J. Mater. Res. {\bf 5},
2356( 1990)

\bibitem{Jeng}D. G. Jeng, H. S. Tuan, R. F. Salat and G. J. Fricano,
Appl. Phys. Lett. {\bf 56}, 1968(1990) 
\bibitem{Yugo}S. Yugo, T. Kanai, T. Kimura and T. Muto, Appl. Phys.
Lett. {\bf 58}, 1036(1991)

\bibitem{Stoner1}B. R. Stoner, G.-H. M. Ma, S. D. Wolter and J. T. Glass,
Phys. Rev. {\bf B45}, 11067(1992) 

\bibitem{Jiang}X. Jiang, K. Schiffmann, A. Westphal and C.-P. Klages,
Appl. Phys. Lett. {\bf 63}, 1203(1993) 

\bibitem{Zhu}W. Zhu, F. R. Sivazlian, B. R. Stoner and J. T. Glass, J.
Mater. Res. {\bf 10}, 425(1995) 

\bibitem{Chen1}Q. Chen and Z. Lin,  Appl. Phys. Lett. {\bf 67}, 1853(1995) 
\bibitem{Chen2}Q. Chen and Z. Lin, Appl. Phys. Lett. {\bf 68}, 2450(1996)
\bibitem{Chen3}Q. Chen and Z. Lin, J. Appl. Phys. {\bf 80}, 797(1996) 
\bibitem{Park}S. S. Park and J. Y. Lee, J. Appl. Phys. {\bf 69},
2618(1991)
\bibitem{Chen4}Qijin Chen, {\it M.S. thesis\/}, Institute of Physics,
Chinese Academy of Sciences, 1995

\bibitem{Lee}S. T. Lee, Y. W. Lam, Z. Lin, Y. Chen and Q. Chen, Phys.
Rev. {\bf B55}, 15937(1997)
\bibitem{Chen5}Q. Chen and Z. Lin, J. Mater. Res. {\bf 10}, 2685(1995)
\bibitem{Spitsyn}B. V. Spitsyn, L. L. Bouilov and B. V. Deryagin, J.
Cryst. Growth {\bf 52}, 219(1981)
\bibitem{Frenklach}M. Frenklach and H. Wang, Phys. Rev. {\bf B43},
1520(1991)
\bibitem{Spear}K. E. Spear, J. Am. Ceram. Soc. {\bf 72}, 171(1989)
\bibitem{Sun1}B. Sun, X. Zhang, Q. Zhang and Z. Lin, J. Appl. Phys.
{\bf 73}, 4614(1993)
\bibitem{size}The radii of hydrogen atoms and and carbon atoms 
differ from different sources. Also it depends on how they are determined.
The covalent radii can differ from the Van der Waals radii by a factor
of two. We take the average for estimate purpose. 
For hydrogen atoms and carbon atoms, they are 0.75\AA\ and 1.3\AA,
respectively. Accordingly, the radii of H$_2$ and $\mathrm{CH_x}$ are
estimated to be 1.0\AA\
and 1.6\AA, respectively.
\bibitem{sqrt2}Due to
different average velocities of hydrogen atoms and of hydrocarbon species,
the factor $\sqrt{2}$ should also be modified accordingly. However, this
is less important in comparison with the size of the gaseous species.
\bibitem{Setaka}N. Setaka, J. Mater. Res. {\bf 4}, 664(1989)
\bibitem{Katoh}M. Katoh, M. Aoki and H. Kawarada, Jpn. J. Appl. Phys.
{\bf 33}, L194(1994)
\bibitem{Stoner2}B. R. Stoner and J. T. Glass, Appl. Phys. Lett. {\bf 60},
698(1992)
\bibitem{Sun2}B. Sun, X. Zhang and Z. Lin, Phys. Rev. {\bf B47},
9816(1993); 
   B. Sun, X. Zhang, Q. Zhang and Z. Lin, Appl. Phys. Lett. {\bf 62},
31(1993)
\bibitem{Chen6}Q. Chen, Y. Chen, J. Yang and Z. Lin, Thin Solid Films
{\bf 274}, 160(1996)
\end{references}
\end{document}